%% file: mnf2009arc.tex
\documentclass{arc}

\gridframe{N}
\jno{}
\usepackage{modamsthm,modnatbib}
\usepackage{latexsym,amssymb,lastpage}
\usepackage{graphicx,amsfonts}
\usepackage{times,mathptmx,bm,amsmath}
\usepackage{dcolumn}
\usepackage[latin1]{inputenc}
\usepackage[absolute]{textpos}

\usepackage{fancyhdr}

\input standard.tex

\renewcommand{\theequation}{\thesection.\arabic{equation}}
\numberwithin{equation}{section}

\newcommand{\new}[1]{{ #1}}

\begin{document}
%\begin{textblock}{8}(13.5,0.5)
%{\it 2nd Micro and Nano Flows Conference}\\
%{\it \hspace*{0.5mm} \mbox{West London, UK, 1-2 September 2009}}
%\end{textblock}
%\rhead[]{
%{\it 2nd Micro and Nano Flows Conference}\\
%{\it West London, UK, 1-2 September 2009}
%}
%\title{Calibration of lubrication force measurements by lattice Boltzmann simulations}
\title{Lattice Boltzmann simulations of liquid film drainage between smooth surfaces}
\author{{\sc Christian Kunert$^1$} and {\sc Jens Harting$^{2,1}$}\\[2pt] 
$^1$Institute for Computational Physics, University of Stuttgart,\\[6pt] 
Pfaffenwaldring 27, D-70569 Stuttgart, Germany\\[6pt]
$^2$Department of Applied Physics,\\[6pt] 
TU Eindhoven, Den Dolech 2, NL-5600MB Eindhoven, The Netherlands\\[6pt]
} 
\maketitle
%\affiliation{* Corresponding author: Email: j.harting@tue.nl\\
%1: Institute for Computational Physics, University of Stuttgart, Pfaffenwaldring 27, D-70569 Stuttgart, Germany\\
%2: Department of Applied Physics, TU Eindhoven, Den Dolech 2, NL-5600MB Eindhoven, The Netherlands}

%\author{Jens Harting}
%\affiliation{Department of Applied Physics, TU Eindhoven, Den Dolech 2, NL-5600MB Eindhoven, The Netherlands}
%\affiliation{Institute for Computational Physics, University of Stuttgart, Pfaffenwaldring 27, D-70569 Stuttgart, Germany}
\begin{abstract}
{\new{ Exploring the hydrodynamic boundary of a surface by approaching a
colloidal sphere and measuring the occurring drag force is a common
experimental technique. However, numerous parameters like the wettability
and surface roughness influence the result. In experiments these cannot be
separated easily.  For a deeper understanding of such surface effects a
tool is required that predicts the influence of different surface
properties. In this paper we present computer simulations based on the
lattice Boltzmann method of a sphere submerged in a Newtonian liquid.  We
show that our method is able to reproduce the theoretical predictions for
flat and noninteracting surfaces.  In order to provide high precision
simulation results the influence of finite size effects has to be well
controlled.  Therefore we study the influence of the required system size
and resolution of the sphere and demonstrate that even moderate
computing resources allow the error to be kept below 1\%.}}
{lattice Boltzmann, microfluidics, lubrication force}
\end{abstract}
%\thispagestyle{fancy} 
%\maketitle
%\makebox[0.3\linewidth][100,100]{XXXXXXXXXXXXXXXXXXXXXXXXXXXXXXXXXXXXXXXXXXXXXXXXXXXXXXXXXXXXXXX}
\section{Introduction}
\new{The miniaturization of technical devices down to
submicrometric sizes has led to the development of so-called
microelectro-mechanical systems (MEMS) which are now commonly applied for
chemical, biological and technical applications (see \cite{bib:tabeling-book}). A
wide variety of microfluidic systems was developed including gas
chromatography systems, electrophoretic separation systems, micromixers, DNA
amplifiers, and chemical reactors. Further,
microfluidic experiments were used to answer fundamental questions in physics including the
behavior of single molecules or particles in fluid flow or the validity of the
no-slip boundary condition (see
\cite{bib:tabeling-book,bib:lauga-brenner-stone-05}).
A violation of the latter in sub-micron sized geometries was found in very well controlled experiments in
recent years.}
Since then, mostly
experimental (\cite{bib:lauga-brenner-stone-05,craig-neto-01,bib:tretheway-meinhart-04,cheng-giordano-02,choi-westin-breuer-03,baudry-charlaix-01,bib:cottin-bizone-etall-02,vinogradova-yakubov-03,bib:churaev-sobolev-somov-84}),
but also theoretical \new{ studies} like \cite{vinogradova-95,bib:degennes-02,bib:bazant-vinogradova-2008}, as well
as computer
simulations (see \cite{succi02,bib:barrat-bocquet-99,bib:cieplak-koplik-banavar-01,bib:thompson-troian-1997})
have been performed to improve our understanding of \new{ slippage}. The
topic is of fundamental interest because it has practical consequences in
the physical and engineering sciences as well as for medical and
industrial applications. Interestingly, also for gas flows, often a slip
length much larger than expected from classical theory can be observed.
Extensive reviews of the slip phenomenon have recently been published by
%Lauga et al.
\cite{bib:lauga-brenner-stone-05},  %Neto et al.
\cite{neto-etal-05}\new{
and ~\cite{bib:bocquet-barrat-2007}}.

Boundary slip is typically quantified by the slip length $b$. This concept
was proposed by \cite{bib:Navier}. He introduced a boundary condition where the
fluid velocity at a surface is proportional to the shear rate at the surface
(at $x=x_0$), i.e.
\begin{equation}
v_z(x_0)=b\frac{\partial v_z(x)}{\partial x}.
\end{equation}
In other words, the slip length $b$ can be defined as the distance
from the surface where the relative flow velocity vanishes.

\begin{figure}[h]
\centerline{
\includegraphics[width=0.45 \linewidth,angle=0]{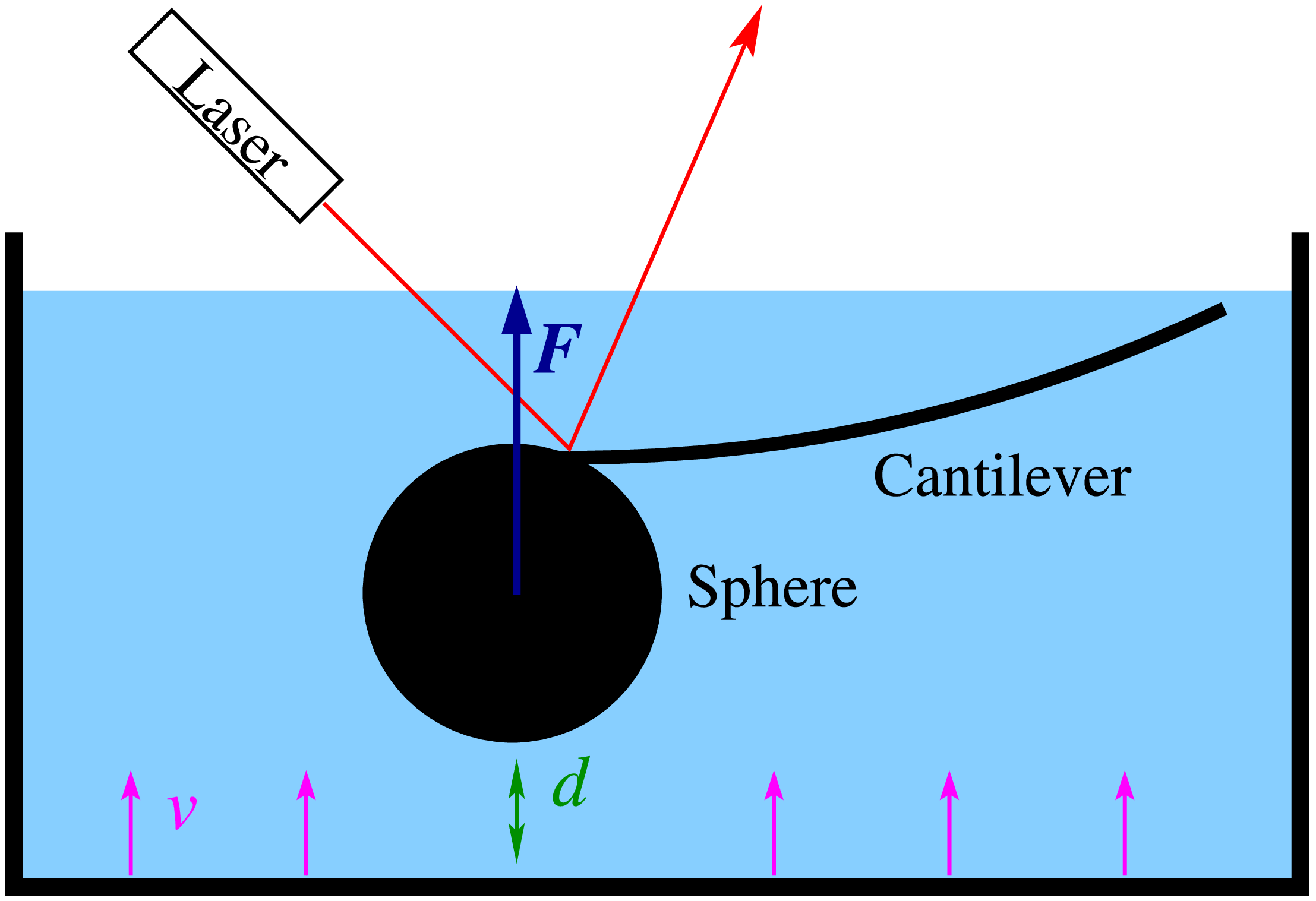}}
\caption{\label{fig:afm-shema}
A sketch of a slip measurement based on a modified atomic force microscope
(AFM). For technical reasons the surface is moved with the velocity $v$
towards the sphere which is attached to the cantilever of the AFM. The laser is
used to measure the bending of the cantilever which can be related to the drag
force $F$.}
\end{figure}

\new{
The experimental investigation of slip can be based on different
setups.
Early reports of slippage measured the mass flow $Q$ through a pipe with radius $R$
with a controlled pressure gradient and compare it to the theoretical    
values for Poiseuille flow. The slip length $b$ can then be derived from the ratio 
between the measured and the theoretical mass flow as
\begin{equation}
\frac{Q_{\beta}}{Q_{\rm Poiseuille}}=1+\frac{4b}{R}.
\label{eq:massflow}
\end{equation}
\cite{schnell-56} used such a method to measure the slip length in coated
glass capillaries  and found $b=1-10\mu {\rm m}$. Other groups found only
a slip length of $b=10-70 {\rm nm}$ (see \cite{bib:lauga-brenner-stone-05}
and \cite{neto-etal-05}).  The problem is that in such experiments a
deviation of the capillary radius $R$  leads automatically to a measured
slip. With other words one can not distinguish between a shift in the
boundary position or an actual slip phenomenon. 

Other experimental methods follow the flow field by introducing tracer
particles into the flow. The basic assumption off all these methods is
that the tracers do have the same velocity as the flow, and that they do
not disturb the flow. An example for such a method is the double focus
cross correlation ( \cite{lumma-etal-03}).  By having two laser beams at a
fixed distance it is possible to record every particle that crosses one of
the focal points of the beams.  The technique rests in the premise that
only a small number of labeled particles are simultaneously located in an
effective focal volume of the order of $10^{-15}{\rm l}$. Therefore the
time cross-correlation can be used to determine the average time a
particle needs to cross the second focus after it crosses the first one.
Since the focus of the laser beams can be located very precisely one has
an accurate measurement of the flow velocity at a given point.
\cite{pit-hervert-leger-00} have applied the cross correlation method for
hexadecane flowing along a hydrocarbon lyophobic smooth surface and found
a slip length of $400 {\rm nm}$. \cite{vinogradova-etal-08} refined the
method and determined a slip length for water and
NaCl aqueous solutions of less than $100 {\rm nm}$ for a
hydrophobic polymer channel which is independent on the shear rate and the
salt-concentration. For hydrophilic surfaces no measurable slip was
detected.

A second example of such a particle based method is micro particle image
velocimetry (micro PIV). The method is very similar to the cross
correlation method, but instead of just taking into account single events
of a particle crossing the focus of a laser beam, one takes a series of
pictures and correlates them to each other.  For macroscopic phenomena
this technique is typically realized by a CCD video camera but on the
microscopic level, more sophisticated methods are needed. This includes
the use of two lasers with different color to illuminate the tracers.
Further, the optical setup is crucial, so that the focal plane of the
optical set up is in the right position.
\cite{tretheway-meinhart-02,tretheway-meinhart-04} applied micro PIV for
water in
hydrophobic glass channels and measured slip lengths of up to $1{\rm\mu
m}$. However \cite{bib:joseph-tabeling-05} found slip lengths for water of less than
$100 {\rm nm}$, stating that this is the minimal resolution of the method,
i.e., that it is doubtful whether there is any slip.

Both methods give the average velocity distribution of the tracer
particles in the flow field. This measured flow profile can then be fitted
by a theoretical flow profile. For a Poiseuille flow, driven by the
pressure gradient $\frac{\partial p}{\partial z}$, with the viscosity
$\mu$ the channel width $d$ and the slip length $b$ on both sides of the
channel, the flow profile in $x$ direction reads as
\begin{equation}
v(x)=\frac{1}{2\mu}\frac{\partial p}{\partial z}\left[d^2-x^2+2dbx\right].
\label{eq:flowprofile}
\end{equation}
As with the simple mass flow measurements reported on above a principle
problem of these tracer based methods is that it is not possible to distinguish between a 
shift in the boundary position and intrinsic slip, i.e. in Eq.~\ref{eq:flowprofile} the 
slip length $b$ and the (effective) channel width $d$ cannot be decoupled
without a variation of $d$.
Another drawback is in the basic assumption of the tracer models, namely that the flow 
velocity is equal and undisturbed by the particles. However on the micro scale the 
particle-particle interaction and particle-wall interaction might become an issue. 
Van der Waals and electrostatic forces between the fluid or surface and the particle can lead to
a significant change of the particle trajectory. 

Another class of experimental methods to determine slip is based on the
measurement of the drag between two surfaces or the force on a particle
moving towards a boundary.
} 
Very popular is the modification of an atomic force microscope (AFM) by
adding a silicon sphere to the tip of the cantilever. A sketch of such an
experiment is shown in Fig.~\ref{fig:afm-shema}. While moving the surface
towards the sphere, the drag force $F$ can be measured with a high precision. 
\new{ Since the typical distance $d$ between the sphere and the surface as
well as the approaching velocity ${\bf v}$ are 
small the simple Reynolds theory for the lubrication force
\begin{equation}
{\bf F}_{\rm Re}=6 \pi \mu {\bf v} R^2/d,
\label{eq:lubrication1}
\end{equation}
can be applied, where $\mu$ is the dynamic viscosity of the
fluid and $R$ the radius of the sphere.
Eq.~\ref{eq:lubrication1} is valid for a no-slip sphere approaching a no-slip surface.
%It is
%possible 
To measure the amount of slip at the wall, $b$, the drag force
is compared with its theoretical value for a slip surface, as reported by
\cite{vinogradova-yakubov-03,bib:vinogradova-96,vinogradova-95}.
A correction
$f^*$ is applied to Eq.~\ref{eq:lubrication1} that takes into account the surface properties
\begin{equation}
\label{eq:fstern}
F_{b}=f^*F(d).
\end{equation}
In case of a surface with the slip length $b$ and a vanishing slip on the
surface of the sphere, the correction $f^*$ is given by~\cite{vinogradova-95} as
\begin{equation}
\label{model2}
   f^{*} = \frac{1}{4} \left( 1 + \frac{3 d}{2 b}\left[ \left( 1 +
\frac{d}{4 b} \right) \ln \left( 1 + \frac{4 b}{d} \right) - 1
\right]\right).
\end{equation}
There are some possibly problematic limitations of this setup. Those
include the hydrodynamic influence of the cantilever, a twist of the
cantilever or uncontrolled roughness on the sphere or the approached
surface.  Further, the exact velocity of the sphere is hard to control
since the drag force and the bending of the cantilever can lead to an
acceleration of the sphere. This contributes to a deviation from the ideal
case and causes sophisticated corrections of the approaching velocity to
be required.  However, if the experimental setup is well controlled, it
allows a very precise measurement and allows one to distinguish between a
shift in the boundary position or slip. 
}

\new{
In general it should be noted that a
significant dispersion
of the slip measurements reported in the literature can be observed even
in similar experimental
systems (see \cite{bib:lauga-brenner-stone-05,neto-etal-05}). 
This is due to the large amount of unknown or uncontrolled parameters. 
}
For example,
observed slip lengths vary between a few
nanometres as reported by \cite{bib:churaev-sobolev-somov-84} and
micrometers as shown in \cite{bib:tretheway-meinhart-04} and while some authors like ~\cite{choi-westin-breuer-03,bib:zhu-granick-01,craig-neto-01} find a
dependence of the slip on the flow
velocity, others like \cite{cheng-giordano-02,bib:tretheway-meinhart-04}
do not.

The large variety of different experimental results to some extent has its
origin in surface-fluid interactions. Their properties and thus their influence
on the experimental results are often unknown and difficult to quantify. In
addition there are many influences that lead to the same effect in a given
experimental setup as intrinsic boundary slip -- as for example a fluid layer
with lower viscosity than the bulk viscosity near the boundary. Unless one is
able to resolve the properties of this boundary layer it cannot be
distinguished from true or intrinsic slip. Such effects can be categorized as
apparent slip.
 
In the literature a large variety of different effects leading to
an apparent slip can be found. However, detailed experimental studies are
often difficult or even impossible. Here computer simulations can be utilized to
predict the influence of effects like surface wettability or
roughness. Most recent simulations apply molecular dynamics and report
increasing slip with decreasing liquid density (see \cite{bib:thompson-robbins-1990})
or liquid-solid
interactions (see \cite{bib:cieplak-koplik-banavar-01,bib:nagayama-cheng-2004}),
while slip decreases with increasing pressure (see \cite{bib:barrat-bocquet-99}).
These simulations are usually limited to some tens of thousands of particles,
length scales of the order of nanometres and time scales of the order of
nanoseconds. Also, shear rates are usually orders of magnitude higher than in
any experiment (see \cite{bib:lauga-brenner-stone-05}). Due to the small accessible
time and length scales of molecular dynamics simulations, mesoscopic simulation
methods like the lattice Boltzmann method are highly applicable for the
simulation of microfluidic experiments. For a simple flow setup like Poiseuille
or Couette flow several investigations of slip models have been 
published by \cite{bib:nie-doolen-chen,succi02,bib:sbragaglia-etal-06,bib:jens-kunert-herrmann:2005,bib:jens-jari:2008},
but investigations for more complex flows are rare.

\new{In this paper lattice Boltzmann simulations of AFM based 
experiments are presented. We focus on demonstrating that our method is able to
reproduce the theoretical prediction for a simple no-slip case and investigate
its limits. It is then possible to use the results to investigate
the influence of different parameters. Such parameters can include an interaction between
the sphere and the boundary, surface roughness or hydrophobicity. While
the case of roughness is investigated in
\cite{bib:jens-kunert-vinogradova:2010}, further studies will be
shown in future publications. Microscopic origins of slip caused by the
molecular details of the fluid and surface cannot be treated with a
continuum method and are beyond the scope of the current paper.} 

The remainder of this paper is arranged as follows: after this introduction we
describe the theoretical background and our simulation method. Then we show
that the method is able to reproduce the theoretical predictions with great
accuracy. Further, we investigate the influence of finite size effects. We
determine the minimum system size and resolution of the discretization required to
push the finite size effects below an acceptable limit and demonstrate
the limits of the simulation method. 

\section{Theoretical background}
\new{In this section the commonly used theory is summarized.
The Reynolds lubrication force is given by Eq.\ref{eq:lubrication1}.}
For larger distances $d$ between the surface of the
sphere and the approached boundary this force does not converge towards the
Stokes drag force ${\bf F}_{\rm St}=6 \pi \mu {\bf v} R$ for a sphere moving
freely in a fluid. Therefore this simple equation fails in cases of
larger separations $d$, where the Stokes force is not
sufficiently small to be neglected. The system can be described accurately by
the theory of~\cite{Maude:1961} and~\cite{Brenner:1961}. The base of the theory is a solution for
two spheres approaching each other with the same rate. By transforming the
coordinates, applying symmetry arguments and setting the radius of one of the
spheres to infinity one arrives at a fast converging sum for the drag force
acting on the sphere:
\begin{equation}
\label{eq:maude}
{\bf F}_{\rm Ma}=6\pi \mu {\bf v} R \lambda_1,
\end{equation}
with
\begin{eqnarray*}
\label{eq:maudea}
&\!\!\!\!\!\!\!\!\!\!\!\lambda_1=-\frac{1}{3} \sinh \xi  \\
&\!\!\!\!\!\!\!\!\!\!\!\times\left(\sum^{\infty}_{n=1} \frac{n(n+1)\left [8e^{(2n+1)\xi}+2(2n+3)(2n-1)\right ]}{(2n-1)(2n+3)[4\sinh^2(n+\frac{1}{2})\xi-(2n+1)^2\sinh^2\xi]}\right.\\
&\!\!\!\!\!\!\!\left.-\sum^{\infty}_{n=1}
\frac{n(n+1)\left [ (2n+1)(2n-1)e^{2\xi}-(2n+1)(2n+3)e^{-2\xi} \right ]}{(2n-1)(2n+3)[4
\sinh^2(n+\frac{1}{2})\xi-(2n+1)^2\sinh^2\xi]}\right),
\end{eqnarray*}
where $\cosh \xi=(d-R)/R$. 
The term given by $\lambda_1$ cannot be treated analytically. Thus, we
evaluate $\lambda_1$ numerically with a convergence of $10^{-10}$.
A more practical approximation of 
(\ref{eq:maude}) is given in the same paper by~\cite{Maude:1961}:
\begin{equation}
\label{eq:forceap}
F(d)=6\pi\mu R v \left (\frac{9}{8}\frac{R}{d}+1 \right )
%\lambda_1 \approx 1/\left(1-\frac{9}{8}\frac{r}{(d+\frac{9}{8}r)}\right )
\end{equation}
Here one can easily see that the force converges towards the Stokes force for an infinite distance $d$ and
towards the Reynolds lubrication (\ref{eq:lubrication1}) for small separations $d$.  

\new{As stated above, to measure the slip length $b$ experiments apply a correction
$f^*$ that takes into account the surface properties so that
$F_{b}=f^*F(d)$.
%\begin{equation}
%\label{eq:fstern1}
%F_{b}=f^*F(d)
%\end{equation}
%In case of a surface with the slip length $b$ and a vanishing slip on the
%surface of the sphere, the correction $f^*$ is given by~\cite{vinogradova-95}
%\begin{equation}
%\label{model2}
%   f^{*} = \frac{1}{4} \left( 1 + \frac{3 d}{2 b}\left[ \left( 1 +
%\frac{d}{4 b} \right) \ln \left( 1 + \frac{4 b}{d} \right) - 1
%\right]\right).
%\end{equation}
The correction function $f^*$ is given by Eq.~\ref{model2}.
This equation is valid for a perfectly flat surface with finite slip, but it
does not allow one to distinguish between slip and other effects like surface
roughness. Therefore it is of importance to perform computer simulations which
have the advantage that all relevant parameters can be changed independently
without modifying anything else in the setup. Thus, the influence of every
single modification can be studied in order to present estimates of the
influence on the measured slip lengths. The first step in this process is to
validate the simulation method and to understand its merits and flaws. In
general most computer simulations suffer from the fact that only a small system
can be described and that one is usually not able to simulate the whole
experimental system in full detail. For this reason it is mandatory to
understand which resolution of the problem is required to keep finite size
effects under control and to cover the important physics correctly.}

\begin{figure}[h]
\centerline{
\includegraphics[width=0.45\linewidth,angle=0]{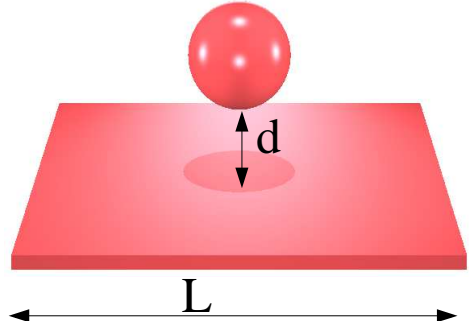}}
\caption{\label{fig:system}
A sketch of the simulated system. The distance between the surface and the
sphere is $d$ and the system length is $L$. The surrounding fluid is not shown. 
}
\end{figure}
In Fig.~\ref{fig:system} a sketch of the simulated system is shown.  A sphere
embedded in a surrounding fluid is simulated. While the sphere approaches a
surface the force acting on it is recorded.  We perform simulations with
different sphere radius $R$ and different system length $L$ to investigate
finite size effects. For simplicity we set the $x$ and $z$ dimension to the
same value $L$ and keep the propagation dimension $y$ constant at $512$ lattice
units. A typical approach to limit finite size effects is to use periodic
boundary conditions. However in such a system the sphere then interacts with
its periodic images. \cite{hasimoto-59} gives a theoretical solution for the drag
force of a sphere in a periodic array as it appears if all
boundaries are periodic:
\begin{equation}
\label{eq:hasimoto}
{\bf F}_{\rm Ha}=\frac{\bf F_{\rm St}}{1-2.83a+4.19a^3-27.4a^7+O(a^9)}
\end{equation}
Here, $a=R/L$ is the ratio between the radius of the sphere $R$ and the
system length $L$. The case we are dealing with in this paper is different
and more complex due to a broken symmetry caused by the approached
surface.  For the approximation given by (\ref{eq:hasimoto}) the main
contribution of the periodic interaction is between the periodic images in
front and behind the sphere. Due to the rigid boundary, these are not
present in our case. 

Besides a finite simulated volume most simulation methods utilize a finite
discretization of the simulated objects, i.e. the sphere in our case.
This means that the finite size and the resolution influence the result of
a simulation. However, it is usually possible to limit the influence of
finite size effects and the loss of accuracy by discretization if those
errors are known and taken into account properly. Therefore, we study the
finite size effects in a simulation of a sphere in a periodic system
approaching a rigid no slip boundary and investigate how different
resolutions of the sphere influence the force acting on it. 

\section{Simulation Method}
The simulation method used to study microfluidic devices has to be chosen
carefully. While Navier-Stokes solvers are able to cover most problems in
fluid dynamics, they lack the possibility to include the influence of
molecular interactions as needed to model boundary slip. Molecular
dynamics simulations (MD) are the best choice to simulate the fluid-wall
interaction, but computer power today is not sufficient to simulate
length and time scales necessary to achieve orders of magnitude which
are relevant for experiments. However, boundary slip with a slip length
$b$ of the order of many molecular diameters $\sigma$ has been studied
with molecular dynamics simulations by various
authors like \cite{bib:thompson-troian-1997,bib:cieplak-koplik-banavar-01,bib:cottin-bizone-etall-04,baudry-charlaix-01}.

In this paper we use the lattice Boltzmann method, where one discretizes the
Boltzmann kinetic equation
\begin{equation}
\label{eq:boltzmann}
\left[\frac{\partial }{\partial t}+{\bf u}\nabla_x+\frac{\hat{\bf F}}{m}\nabla_u
\right] n({\bf x},{\bf u},t)={\bf \Omega}
\end{equation}
on a lattice. $n({\bf x,u},t)$ indicates the probability to find a single
particle with mass $m$ and velocity ${\bf u}$ at the time $t$ and position
${\bf x}$. $\hat{\bf F}$ accounts for external forces. The derivatives
represent simple propagation of a single particle in real and velocity space
whereas the collision operator ${\bf \Omega}$ takes into account molecular
collisions in which a particle changes its momentum due to a collision with
another particle. Further, the collision operator drives the distribution
$n$ towards an equilibrium distribution $n^{eq}$.

In the lattice Boltzmann method time, positions, and velocity space are
discretized on a lattice in the following way. The distribution $n$ is only
present on lattice nodes ${\bf x}_k$. The velocity space is discretized so that
in one discrete timestep $\delta t$ the particles travel with the discrete
velocities ${\bf c}_i$ towards the nearest and next nearest neighbours ${\bf
x}_k+{\bf c}_i \delta t$. Since a large proportion of the distribution stays at
the same lattice node a rest velocity ${\bf c}_0$ is required. ${\bf c}_0$
represents particles not moving to a neighboring site. In short we operate on a
three dimensional grid with 19 velocities ($i=0..18$) which is commonly
referred to as D3Q19. After the streaming of the population density, the
population on each lattice node is relaxed towards an equilibrium such that
mass and momentum are conserved.  It can be shown by a Chapman-Enskog procedure
that such a simulation method reproduces the Navier-Stokes
equation, \cite{bib:succi-01}. In the lattice Boltzmann method the time $t$ is
discretized in time steps $\delta t$, the position ${\bf x}$ is discretized in
units of distance between neighbouring lattice cells, and the velocity ${\bf
u}$ is discretized using the velocity vectors ${\bf c}_i$. These form the
natural lattice units of the method which are used in this paper if not stated
otherwise.
  
The implementation we are using originates from
\cite{bib:verberg-ladd-01}. It applies a so called multi relaxation
time collision operator. The distribution $n$ is transformed via a
transformation matrix $T$ into the space of the moments $m_j$ of the
distribution. A very accessible feature of this approach is that some of the
moments $m_j$ have a physical meaning as for example the density
\begin{equation}
\rho=m_0=\sum_i n_i
\end{equation}
or the momentum in each direction $x,y,z$,
\begin{equation}
m_{1,2,3}={\bf e}_{1,2,3}\sum_i n_i {\bf c}_i.
\end{equation}
${\bf e}_{1,2,3}$ is the unit vector in Cartesian directions.  The moments
$m_j$ relax with an individual rate $S_j$ towards the equilibrium  $m_j^{eq}$.
The equilibrium distribution $m^{eq}$ conserves mass $m_0$ and momentum
$m_{1,2,3}$ and is a discretized version of the Maxwell distribution.  Thus the
lattice Boltzmann equation (\ref{eq:boltzmann}) can be written as
\begin{eqnarray}
\label{eq:boltzmanna}
n({\bf x}_k+{\bf c}_i \delta t,{\bf c}_i,t+\delta t)-n({\bf x}_k,{\bf c}_i,t)\nonumber\\ 
= {T_{i}^{j}}^{-1} S_j[m_j({\bf x}_k,t)-m_j^{eq}].
\end{eqnarray}
The multi relaxation time approach has several advantages compared to
other lattice Boltzmann schemes. These include a higher precision at solid
boundaries and the direct accessibility of the moments which represent
physical properties. The latter can be utilized to easily implement
thermal fluctuations or external and internal
forces (see \cite{bib:duenweg-schiller-ladd:2007}).

A feature of the implementation we are using is the possibility to simulate
particles suspended in fluid. The simulation method is described extensively in
the
literature (see
\cite{bib:verberg-ladd-00a,bib:verberg-ladd-01,bib:duenweg-schiller-ladd:2007,bib:jens-komnik-herrmann:2004}).
Therefore only a brief description is given here. The movement of the particles
is described by a simple molecular dynamics algorithm. However it should be
noted here that we simulate a sphere moving with a constant velocity. Therefore
any forces acting on the sphere do not influence its movement. The
fluid-particle interaction is achieved by the solid-fluid boundary interaction
acting on the surface of the sphere. When the sphere is discretized on the
lattice all lattice sites inside the sphere are marked as boundary nodes with a
moving wall boundary condition. This boundary condition at the solid-fluid
interface is constructed in such a way that there is as much momentum
transferred to the fluid as required for the fluid velocity to match the
boundary velocity ${\bf v}_b$ of the particle. The center of mass velocity of
the particle and the rotation are taken into account. This way the transfered
momentum and thus the hydrodynamic force acting on the sphere are known.
Technically speaking a link bounce back boundary condition is implemented for
the solid nodes together with a momentum transfer term. The link bounce back
implies that the distributions that would move inside the boundary with the
velocity ${\bf c}_i$ are reversed in direction with opposite velocity ${\bf
c}_k$.
\begin{eqnarray}
\!\!\!\!\!\!\!\!\!\!\!\!\!
n({\bf x}+{\bf c}_i\delta t,{\bf c}_i,t+\delta t)&\!\!\!=\!\!\!&n({\bf x}+{\bf c}_i\delta t,{\bf c}_k,t)\!+\!\frac{2a^{{\bf c}_i}\rho {\bf v}_b {\bf c}_i} {c_s^2}\\
n({\bf x},{\bf c}_k,t+\delta t)&\!\!\!=\!\!\!&n({\bf x},{\bf c}_i,t)\!+\!\frac{2a^{{\bf c}_i}\rho {\bf v}_b c_i}{c_s^2}
\label{eq:movingbb}
\end{eqnarray}
Here, $a^{{\bf c}_i}$ are weight factors taking into account the different
lengths of the lattice vectors.

While the center of mass of the sphere moves, new lattice nodes become
part of the particle, while others become fluid. Therefore particles do
not perform a continuous movement but rather small jumps. After each jump
the fluid is out of equilibrium but relaxes very quickly back to the quasi
static state. In order to average out statistical fluctuations imposed by
the discrete movement of the particle, one has to average the recorded
force over several time steps. We choose to average over intervals of
$999$ steps. 

If not stated otherwise the simulation parameters are ${ v}=0.001$, 
$\mu=0.1$ and the radius is varied between $R=4$ and $R=16$.  The approached
boundary is a plain no-slip wall which is realized by a mid grid bounce
back boundary condition.

Along the open sides periodic boundary conditions are applied so that the
sphere can interact with its mirror leading to the to be avoided finite size
effects. As noted in Fig.~\ref{fig:system} the length of the system in $x$ and
$y$ direction is $L$. The size of the simulation volume is varied to explore
the influence of finite size effects.

\section{Results}
In this contribution we vary the system length $L$ and the radius of the sphere
$R$. A major contribution to the finite size effects is the interaction of the
sphere with its periodic image. Therefore a larger system length should reduce
this effect dramatically. However when the hydrodynamic influence of the wall
becomes larger finite size effects become smaller. This can be explained by the
fact that the friction at the boundary suppresses the hydrodynamic interaction
of the particle with its periodic image. Instead the dominant interaction is
between the particle and the surface. It is mandatory for a better
understanding of the system to learn how these finite size effects can be described,
quantified, and controlled. 

\begin{figure}[h]
\centerline{
\includegraphics[width=0.35 \linewidth,angle=270]{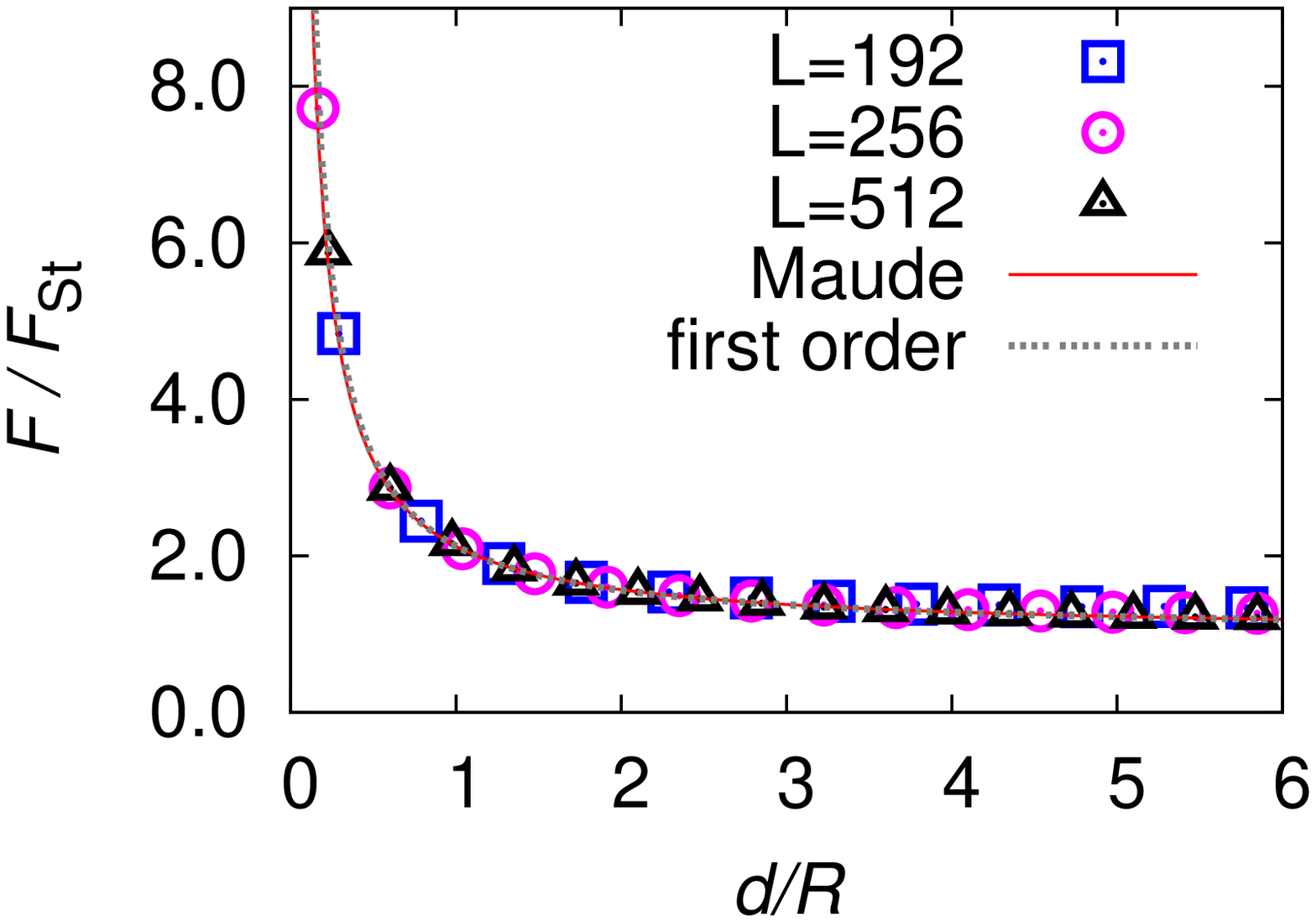}
\includegraphics[width=0.35 \linewidth ,angle=270]{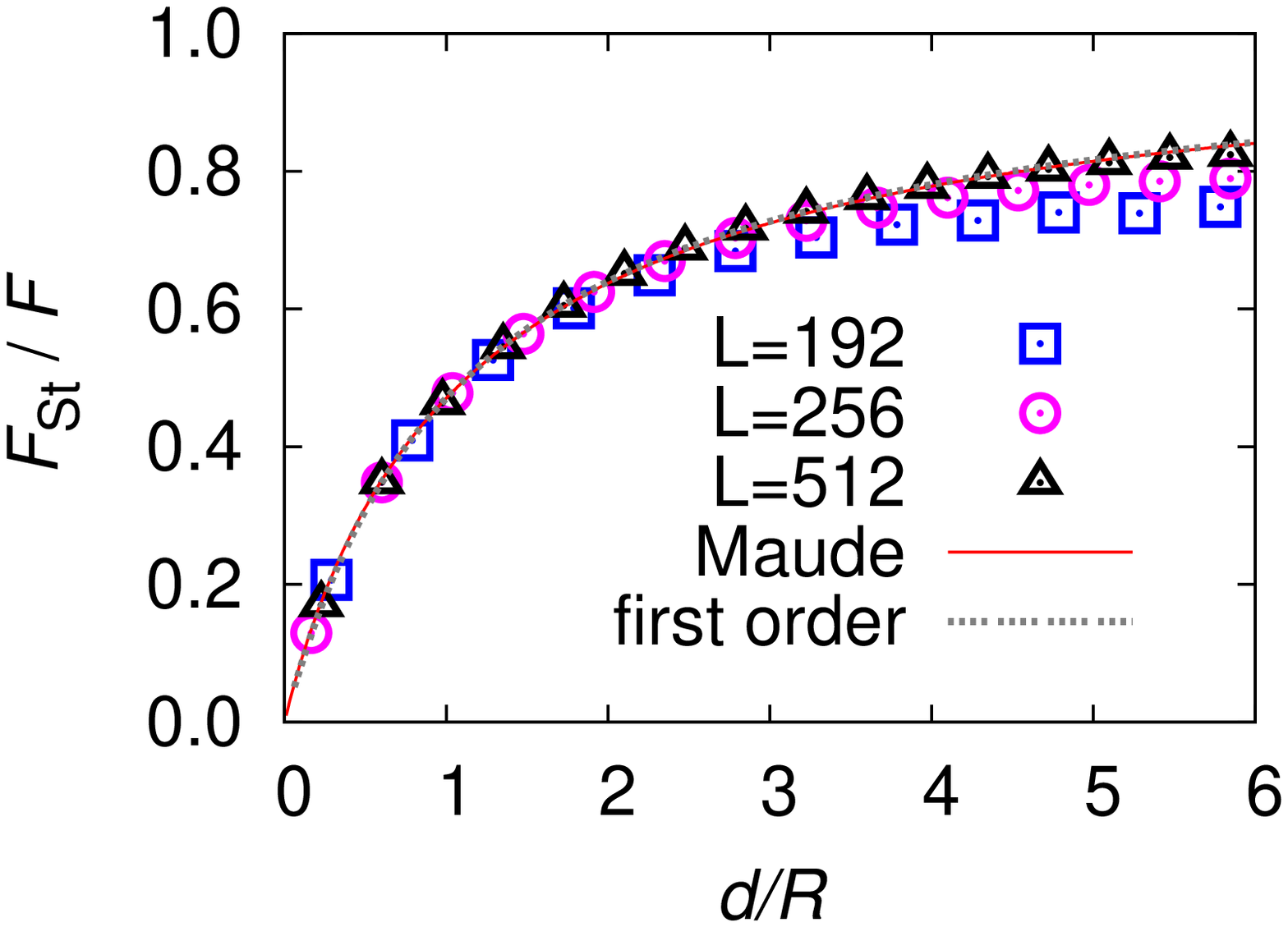}}
\caption{\label{fig:forceR16}
Normalized drag force $\frac{ F}{ F_{St}}$ and the inverted
normalized drag force $\frac{ F_{St}}{ F}$ versus the normalized
distance $d/R$ for different system lengths $L$. The radius of the sphere is
fixed at $R=16$. The deviation for $L=192$ is significant at larger radii, but
can be neglected for small distances. For $L=512$ there is nearly no deviation
from the exact solution of Maude (\ref{eq:maude}). In addition the deviation
of the first order approximation (\ref{eq:forceap}) is below $1 \%$ for
$d>R/2$. 
}
\end{figure}
First we study a system with a constant sphere radius $R=16$ and varying system
length $L$. In Fig.~\ref{fig:forceR16} the drag force ${ F}$ normalized by
the Stokes force ${F}_{\rm St}=6 \pi \mu {v} R$ and the inverse
normalized drag force are plotted. In the inversed case the deviations for the
larger distance $d$ can be seen more clearly. Fig.~\ref{fig:forceR16} shows
that the deviation for the small system $L=192$ close to the wall is very
small, however the force does not converge to the Stokes force ${ F}_{\rm
St}$. Here effects similar to the one reported by Hasimoto (\ref{eq:hasimoto})
appear: for a smaller separation $d$ the force decays with $\frac{1}{d}$ while
it approaches a constant value for large $d$. The constant values should be
given by the Stokes force ${F}_{\rm St}$, but can be larger due to the
interaction with the periodic image. In the $1/F$ plots it can be seen that the
deviation is not a constant offset or factor but rather starts at a critical
value of $d/R$. From there the force quickly starts to approach a constant
value. 

\begin{figure}[h]
\centerline{
\includegraphics[width=0.35 \linewidth,angle=270]{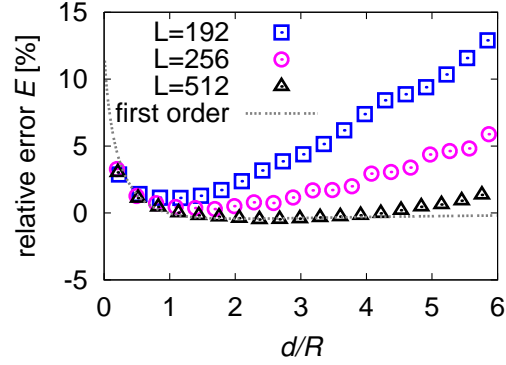}}
\caption{\label{fig:errorn}
Symbols denote the relative error $E=\frac{F-F_{Maude}}{F_{Maude}}$ in \%
versus the normalized distance $d/R$ for different system sizes $L$. The line
shows the first order approximation. As expected the error close to the wall
deviates due to the discretization of the small distance $d$. The error becomes
larger for large $d$ due to the influence of the periodic image.
}
\end{figure}

In  Fig.~\ref{fig:errorn} the relative error $E=\frac{F-F_{Maude}}{F}$ is
plotted for different system sizes $L$ and a constant radius $R$. The error for
the largest system $L=512$ in Fig~\ref{fig:errorn} is constantly below $1\%$
for larger distances. At distances less than $d<R/2$ the error rises due to the
insufficient resolution of the fluid filled volume between the surface of the
sphere and the boundary. Another possible effect is the fact that the sphere
rather jumps over the lattice than performing a continuous movement.
Additionally it can be seen that for distances less than $d=R$ the error for
the different system sizes $L$ collapses. The reason is that for smaller
distances the lubrication effect which is independent of the system length $L$
dominates the free flow and therefore suppresses finite size effects due to the
periodic image. The deviation that can be seen in the plot at the large $d$
has its origin in the transient. Since the fluid is at rest at the start of the
simulation and it takes some time to reach a steady state this can only be
avoided by longer simulations. An interesting fact to point out is that the
deviation between the Maude solution and the first order approximation is below
$1\%$.
 
\begin{figure}[h]
\centerline{
\includegraphics[width=0.35 \linewidth ,angle=270]{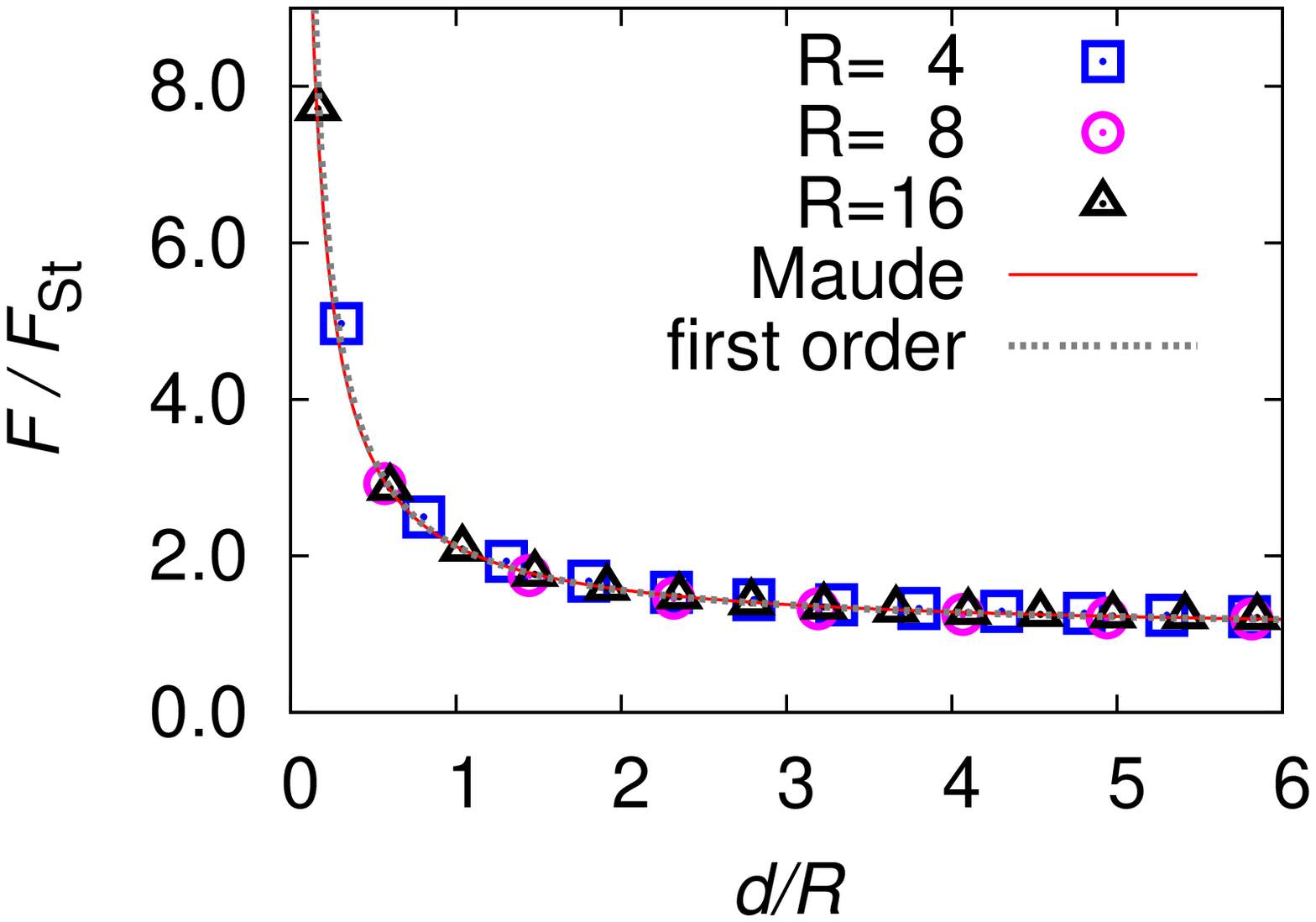}
\includegraphics[width=0.35 \linewidth ,angle=270]{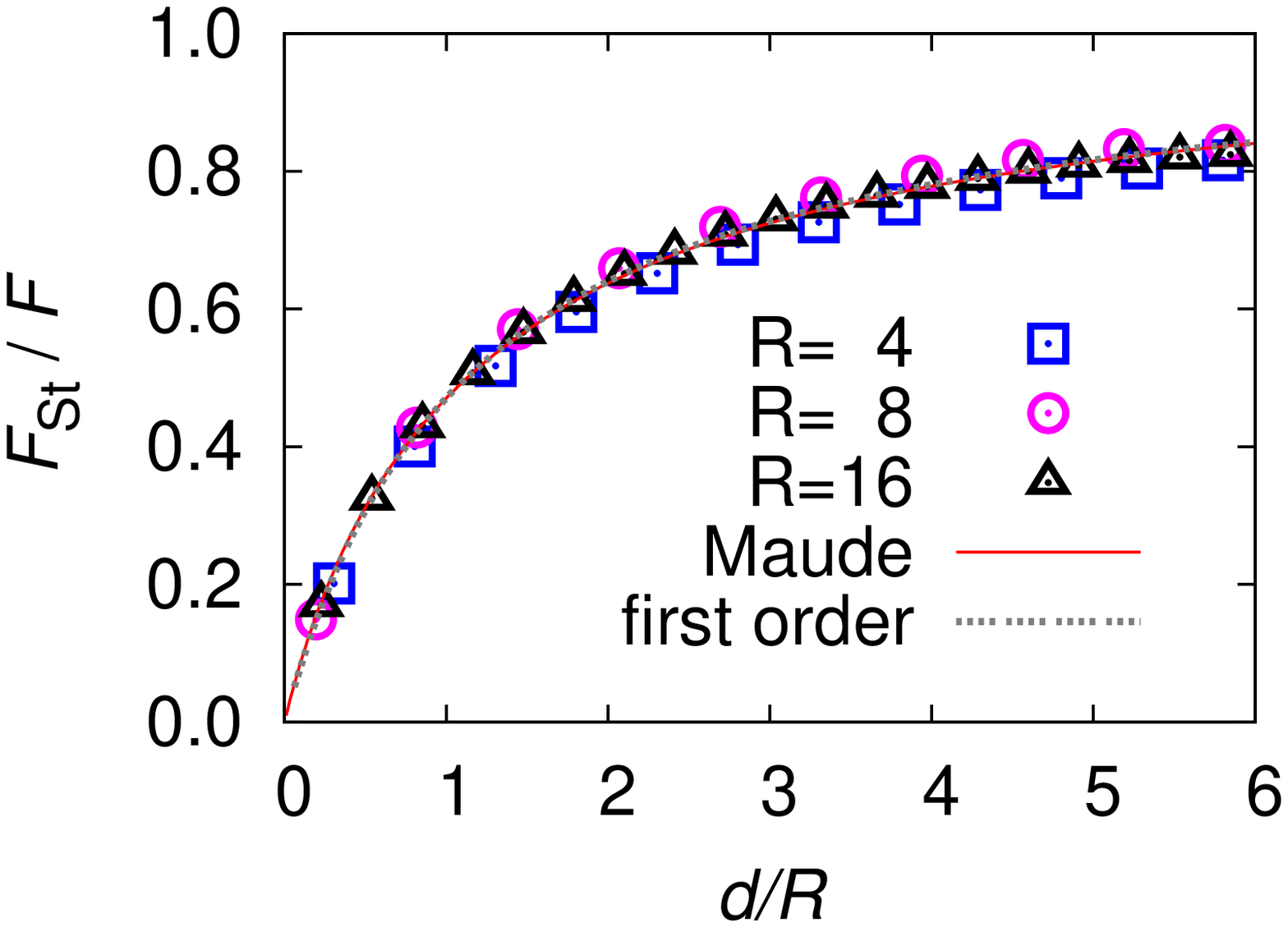}}
\caption{\label{fig:forceR}
Normalized \new{ drag} force $\frac{ F}{ F_{\rm St}}$ and the inverted
normalized \new{ drag} force $\frac{ F_{\rm St}}{ F}$ versus the
normalized distance $d/R$ for different radii $R$ but constant ratio
$R/L=1/32$
}
\end{figure}

Since the sphere is discretized on the lattice it is important to understand if
this discretization has an effect on the {\new drag} force. Therefore we
perform simulations with a radius of $R={4,8,16} $ at a constant ratio
$R/L=1/32$ between the radius and the system length. Fig.~\ref{fig:forceR}
depicts the normalized \new{ drag} force $\frac{ F}{ F_{\rm St}}$ and the
inverted normalized \new{ drag} force $\frac{ F_{\rm St}}{ F}$ versus the
normalized distance $d/R$ for different radii $R$. It can be seen that the
discretization of the sphere has little influence on the measured force. 

\begin{figure}[h]
\centerline{
\includegraphics[width=0.35 \linewidth, angle=270]{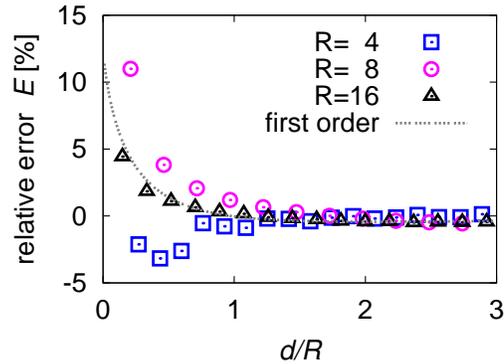}}
\caption{\label{fig:errorR}
Relative error $E=\frac{ F-F_{\rm Maude}}{ F_{\rm Maude}}$ in \% versus the
normalized distance $d/R$ for different radii $L$ at a constant ratio
$R/L=\frac{4}{128}$ (symbols). As expected the error close to the wall
deviates, due to the discretization of the small distance $d$. For $R=4$ the
deviation fluctuates due to the low resolution but for $R=16$ it follows the
first order approximation. The line corresponds to the first order
approximation.
}
\end{figure}

Fig.~\ref{fig:errorR} shows the relative error $E$ for different radii. For all
radii the finite size effects due to the periodic image are negligible since
the ratio between $R/L$ is sufficiently small. The deviation from the Maude
theory for separations $d>R$ are below $2\%$ for all radii. Therefore one has
to concentrate on the small distances $d$ where significant deviations appear.
In our case this distance is better resolved for larger $R$ (note that in the
plot the normalized distance is shown). In addition the resolution of the
sphere is better for larger $R$. For $R=4$ the deviation is more noisy and here
the discretization really has an effect on the drag force ${\bf F}$.
However, even a such roughly discretized sphere reproduces the expected
result surprisingly well. This can probably be explained by averaging out
of discretization artefacts due to the time and space averaging performed
in the simulations.
Additionally there should be three or more lattice sites between the surface of
the sphere and the boundary. If that is not the case the hydrodynamic
interaction is not resolved sufficiently. If the distance between surface and
sphere is smaller than half a lattice spacing the two surfaces merge and the
method fails. Hence it is advantageous to choose a large radius in order to be
able to reduce the relative distance to the boundary (in units of the sphere
radius) or to resolve a possible surface structure.

For $R\ge 8$ the deviations have a regular shape and follow the deviation for
the first order approximation.  The trend to follow the first order
approximation is stronger for $R=16$ but here the noise is reduced further and
all errors seem to be systematic. Therefore the deviation has to be described
as a systematic error of the method that has its origin in the
``jumping-standing'' like movement of the sphere. The first order approximation
is a quasi static approximation and represents the actual simulated movement
more correctly than the theory of Maude.

It should be noted that the finite size effects for $R=16$ and $R/L=1/16$ due
to the interaction with the periodic image are much more significant than the
discretization effect.  By choosing a large simulation volume, a radius $R>8$
and focusing on the force for separations $d<2R$ one can reduce those effects
to a deviation of the measured force from the theoretically predicted value of
less than $1\%$.

\section{Conclusion} 
\new{
Lattice Boltzmann simulations of a high-speed drainage of liquid films
squeezed between a sphere and a smooth no-slip surface have been
presented. We have shown that the solution of Maude (\ref{eq:maude}) can
be reproduced and demonstrated that at a ratio $R/L=1/32$ finite size
effects are below $2\%$ and thus can be neglected near the boundary. We
have also demonstrated that a sphere radius of $R=8$ provides a
sufficiently well resolved representation of the sphere. Based on this
calibration it is possible to investigate the influence of different
surface properties such as roughness and slip on the drag force on AFM
based slip measurements.
}
 
\section*{Acknowledgments}
We like to thank O.I. Vinogradova for fruitful discussions, A.~Ladd for
providing the simulation code and helpful suggestions, the DAAD and the DFG
priority program SPP 1164 for funding. Simulations have been performed at the
Scientific Supercomputing Centre, Karlsruhe and the J\"ulich Supercomputing
Center. 

%\bibliographystyle{agsm}
%\bibliography{main_updated,jens-pub.bib}

\end{document}

%% file: standard.tex
 \newtheoremstyle{theorem}{6pt}{6pt}{\rm}{}{\sffamily}{ }{ }{}
 \theoremstyle{theorem}

  \newtheoremstyle{thm}{6pt}{6pt}{\rm}{}{\sffamily}{ }{ }{}
 \theoremstyle{thm}

 \newtheoremstyle{lemma}{6pt}{6pt}{\rm}{}{\sffamily}{ }{ }{}
 \theoremstyle{lemma}
 
 \newtheoremstyle{lem}{6pt}{6pt}{\rm}{}{\sffamily}{ }{ }{}
 \theoremstyle{lem}

\newtheoremstyle{case}{6pt}{6pt}{\rm}{}{}{. }{ }{}
 \theoremstyle{case}

 \newtheoremstyle{statement}{6pt}{6pt}{\rm}{}{\sffamily}{ }{ }{}
\theoremstyle{statement}

 \newtheoremstyle{corollary}{6pt}{6pt}{\rm}{}{\sffamily}{ }{ }{}
 \theoremstyle{corollary}

  \newtheoremstyle{defi}{6pt}{6pt}{\rm}{}{\sffamily}{ }{ }{}
 \theoremstyle{defi}

  \newtheoremstyle{cor}{6pt}{6pt}{\rm}{}{\sffamily}{ }{ }{}
 \theoremstyle{cor}

\newtheoremstyle{example}{6pt}{6pt}{\rm}{}{\sffamily}{ }{ }{}
\theoremstyle{example}

\newtheoremstyle{remark}{6pt}{6pt}{\rm}{}{\sffamily}{ }{ }{}
\theoremstyle{remark}

\newtheoremstyle{approximation}{6pt}{6pt}{\rm}{}{\sffamily}{ }{ }{}
\theoremstyle{approximation}

\newtheoremstyle{scheme}{6pt}{6pt}{\rm}{}{\sffamily}{ }{ }{}
\theoremstyle{scheme}

\newtheoremstyle{Algorithm}{6pt}{6pt}{\rm}{}{\sffamily}{ }{ }{}
\theoremstyle{Algorithm}

 \newtheoremstyle{Remark}{6pt}{6pt}{\rm}{}{\sffamily}{ }{ }{}
 \theoremstyle{Remark}

\newtheoremstyle{Lemma}{6pt}{6pt}{\rm}{}{\sffamily}{ }{ }{}
\theoremstyle{Lemma}

\newtheoremstyle{Assumption}{6pt}{6pt}{\rm}{}{\sffamily}{ }{ }{}
\theoremstyle{Assumption}

\newtheoremstyle{Proposition}{6pt}{6pt}{\rm}{}{\sffamily}{ }{ }{}
\theoremstyle{Proposition}

\newtheoremstyle{prop}{6pt}{6pt}{\rm}{}{\sffamily}{ }{ }{}
\theoremstyle{prop}

\newtheoremstyle{rem}{6pt}{6pt}{\rm}{}{\sffamily}{ }{ }{}
 \theoremstyle{rem}

\newtheoremstyle{hypo}{6pt}{6pt}{\rm}{}{\sffamily}{ }{ }{}
 \theoremstyle{hypo}

  \newtheoremstyle{Step}{6pt}{6pt}{\rm}{}{}{ }{ }{}
 \theoremstyle{Step}

 \newtheoremstyle{lema}{6pt}{6pt}{\rm}{}{\sffamily}{ }{ }{}
 \theoremstyle{lema}